\documentclass[twocolumn]{aastex63}

\received{2022 Oct 14}
\accepted{2022 Dec 29}

\shorttitle{CO(1--0) of TN J0924$-$2201}
\shortauthors{Lee et al.}
\graphicspath{{./}{figures/}}

\begin{document}

\title{Massive molecular gas companions uncovered by VLA CO(1--0) observations of\\ the $z$ = 5.2 radio galaxy TN J0924$-$2201}

\correspondingauthor{Kianhong Lee}
\email{kianhong.lee@nao.ac.jp}

\author[0000-0003-4814-0101]{Kianhong Lee}
\affiliation{Institute of Astronomy, The University of Tokyo,
Mitaka, Tokyo 181-0015, Japan}
\affiliation{National Astronomical Observatory of Japan, 2-21-1 Osawa, Mitaka, Tokyo 181-8588, Japan}

\author[0000-0002-4052-2394]{Kotaro Kohno}
\affiliation{Institute of Astronomy, The University of Tokyo,
Mitaka, Tokyo 181-0015, Japan}
\affiliation{Research Center for the Early Universe, School of Science, The University of Tokyo, 7-3-1 Hongo, Bunkyo-ku, Tokyo 113-0033, Japan}

\author[0000-0001-6469-8725]{Bunyo Hatsukade}
\affiliation{Institute of Astronomy, The University of Tokyo,
Mitaka, Tokyo 181-0015, Japan}

\author[0000-0002-1639-1515]{Fumi Egusa}
\affiliation{Institute of Astronomy, The University of Tokyo,
Mitaka, Tokyo 181-0015, Japan}

\author[0000-0002-4999-9965]{Takuji Yamashita}
\affiliation{National Astronomical Observatory of Japan, 2-21-1 Osawa, Mitaka, Tokyo 181-8588, Japan}
\affiliation{Research Center for Space and Cosmic Evolution, Ehime University, 2-5 Bunkyo-cho, Matsuyama, Ehime 790-8577, Japan}

\author{Malte Schramm}
\affiliation{National Astronomical Observatory of Japan, 2-21-1 Osawa, Mitaka, Tokyo 181-8588, Japan}
\affiliation{Graduate School of Science and Engineering, Saitama University, Shimo-Okubo 255, Sakura-ku, Saitama-shi, Saitama 338-8570, Japan}

\author{Kohei Ichikawa}
\affiliation{Frontier Research Institute for Interdisciplinary Sciences, Tohoku University, Sendai 980-8578, Japan}
\affiliation{Astronomical Institute, Tohoku University, Aramaki, Aoba-ku, Sendai, Miyagi 980-8578, Japan}
\affiliation{Max-Planck-Institut f\"{u}r extraterrestrische Physik (MPE), Giessenbachstrasse 1, D-85748 Garching bei M\"{u}unchen, Germany}

\author{Masatoshi Imanishi}
\affiliation{National Astronomical Observatory of Japan, 2-21-1 Osawa, Mitaka, Tokyo 181-8588, Japan}
\affiliation{Department of Astronomical Science, Graduate University for Advanced Studies (SOKENDAI), Mitaka, Tokyo 181-8588, Japan}

\author{Takuma Izumi}
\affiliation{National Astronomical Observatory of Japan, 2-21-1 Osawa, Mitaka, Tokyo 181-8588, Japan}

\author{Tohru Nagao}
\affiliation{Research Center for Space and Cosmic Evolution, Ehime University, 2-5 Bunkyo-cho, Matsuyama, Ehime 790-8577, Japan}

\author[0000-0002-3531-7863]{Yoshiki Toba}
\affiliation{National Astronomical Observatory of Japan, 2-21-1 Osawa, Mitaka, Tokyo 181-8588, Japan}
\affiliation{Department of Astronomy, Kyoto University, Kitashirakawa-Oiwake-cho, Sakyo-ku, Kyoto 606-8502, Japan}
\affiliation{Research Center for Space and Cosmic Evolution, Ehime University, 2-5 Bunkyo-cho, Matsuyama, Ehime 790-8577, Japan}
\affiliation{Academia Sinica Institute of Astronomy and Astrophysics, 11F of Astronomy-Mathematics Building, AS/NTU, No.1, Section 4, Roosevelt Road, Taipei 10617, Taiwan}

\author{Hideki Umehata}
\affiliation{Institute for Advanced Research, Nagoya University, Furocho, Chikusa, Nagoya 464-8602, Japan}
\affiliation{Department of Physics, Graduate School of Science, Nagoya University, Furocho, Chikusa,
Nagoya 464-8602, Japan}

\begin{abstract}

We present \textit{Karl G. Jansky} Very Large Array (VLA) K-band (19 GHz) observations of the redshifted CO(1--0) line emission toward the radio galaxy TN J0924$-$2201 at $z=5.2$, which is one of the most distant CO-detected radio galaxies. 
With the angular resolution of $\sim2''$, the CO(1--0) line emission is resolved into three clumps, within $\pm500$ km\,s$^{-1}$ relative to its redshift, where is determined by Ly$\alpha$. We find that they locate off-center and 12--33 kpc away from the center of the host galaxy, which has counterparts in $HST$ $i$-band, $Spitzer$/IRAC and ALMA Band-6 (230 GHz; 1.3 mm). 
With the ALMA detection, we estimate $L_{\rm IR}$ and SFR of the host galaxy to be $(9.3\pm1.7)\times10^{11} L_{\odot}$ and $110\pm20$ $M_{\odot}\,\rm yr^{-1}$, respectively.
We also derive the $3\sigma$ upper limit of $M_{\rm H_{2}}<1.3\times10^{10}$ $M_{\odot}$ at the host galaxy.
The detected CO(1--0) line luminosities of three clumps, 
$L'_{\rm CO(1-0)}$ = (3.2--4.7)$\times10^{10}$ $\rm\,K\,km\,s^{-1}pc^{2}$,
indicate the presence of three massive molecular gas reservoirs with $M_{\rm H_{2}}$ = (2.5--3.7)$\times10^{10}$ $M_{\odot}$, 
by assuming the CO-to-H$_{2}$ conversion factor $\alpha_{\rm CO} = 0.8$ $M_{\rm \odot}\rm\,(K\,km\,s^{-1}pc^{2})^{-1}$, 
although the star formation rate (SFR) is not elevated because of the non-detection of ALMA 1.3 mm continuum (SFR $<$ 40 $M_\odot$ yr$^{-1}$). 
From the host galaxy, the nearest molecular gas clump labeled as clump A, is apparently aligning with the radio jet axis, showing the radio-CO alignment. 
The possible origin of these three clumps around TN J0924--2201 can be interpreted as merger, jet-induced metal enrichment and outflow.

\end{abstract}


\section{Introduction} \label{sec:intro}

\begin{figure}[ht!]
\epsscale{1.25}
\plotone{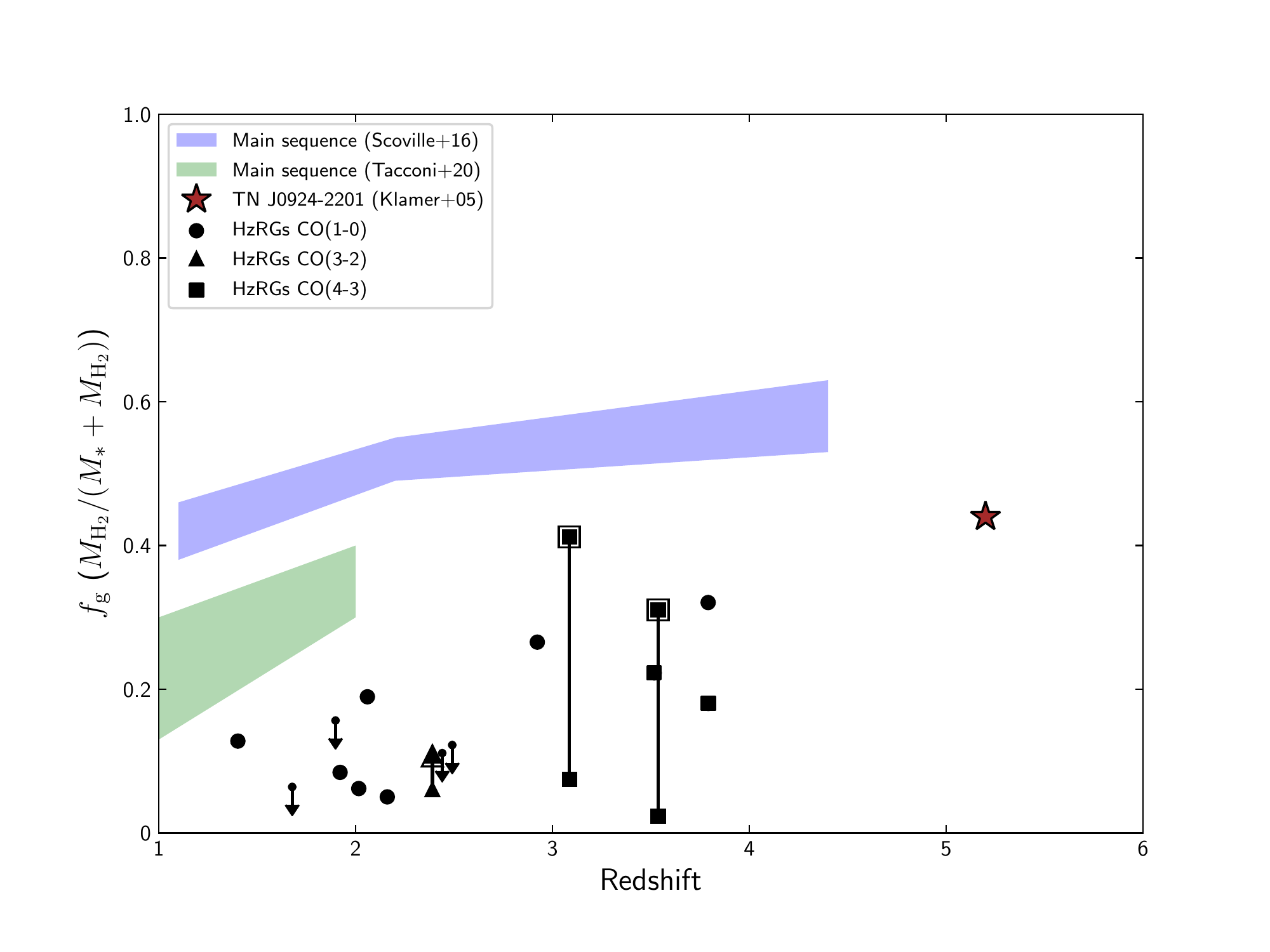}
\caption{The molecular gas fraction $f_{\rm g}$ of HzRGs at different redshifts. Blue and green shaded area show the star-forming main sequence from \citet{Scoville16} and \citet{Tacconi20}, respectively.
Stellar masses of main sequence here are $>10^{11}$ $M_{\rm \odot}$ which is similar to HzRGs. Black symbols show the HzRGs in the literature \citep{Emonts13,Emonts14,Papadopoulos00,Papadopoulos05,Greve04,DeBreuck05,Gullberg16a}. The circles, triangles and squares represent the observations based on CO(1$-$0), CO(3$-$2) and CO(4$-$3), respectively, including $3\sigma$ upper limits. Those with larger open square and triangle symbols are assumed to have $M_{*}=10^{11}M_{\odot}$ when deriving their possible range of $f_{\rm g}$. The vertical line between symbols show the range of $f_{\rm g}$, the lower limits are derived with the upper limits of their $M_{*}$, and the higher values are derived by assuming $M_{*}=10^{11}M_{\odot}$. Stellar masses of HzRGs are from \citet{DeBreuck10}.
The star symbol indicates the $z=5.2$ radio galaxy TN J0924$-$2201 \citep[derived from unresolved CO(1--0);][]{Klamer05}. 
\label{fig:fgvsz}}
\end{figure}

High redshift radio galaxies (HzRGs), are the most massive (stellar mass $M_{*}\gtrsim10^{11}M_{\odot}$) galaxies in 
the early universe, 
with the characteristic powerful radio jets (radio luminosity $L_{\rm500MHz}>10^{27}$ $\rm W\,Hz^{-1}$) at $z\gtrsim2$ \citep[][for a review]{MileyDeBeruck08}.
In contrast to the local quiescent massive galaxies, HzRGs are actively star-forming, lying either on or below the star-forming main sequence \citep{Drouart14,Falkendal19}. Comparing with other populations of galaxies, HzRGs show high star formation efficiencies of $10^{0}$--$10^{2}$ Gyr$^{-1}$, short depletion time of $10^{0}$--$10^{-2}$ Gyr \citep{Man19}, and low molecular gas fraction $f_{g}\equiv M_{\rm H_{2}}/(M_{*}+M_{\rm H_{2}})$, where $M_{\rm H_{2}}$ is molecular gas mass (see Figure~\ref{fig:fgvsz}).

Regarding the molecular gas mass measurements, as CO(1--0) line emission is one of the most convincing tracers of molecular gas, \cite{Emonts13} reported the CO(1--0) detection in a HzRG at $z\sim2$ and \cite{Emonts14} conducted a CO(1--0) survey of $z\sim2$ HzRGs to measure the amount of molecular gas in HzRGs.
They find that HzRGs with stellar masses $M_{*}\sim10^{11}$--$10^{12}$ $M_{\odot}$ have molecular gas masses $M_{\rm H_{2}}\sim10^{10}$--$10^{11}$ $M_{\odot}$. 
At z$\sim$3--4, there are also several studies that have estimated the molecular gas masses of HzRGs by observing CO(1--0), CO(2--1) or CO(4--3) \citep{Papadopoulos00, Papadopoulos05, Greve04, DeBreuck05, Gullberg16a}.
Combing these measurements of molecular gas masses in the literature and the estimation of stellar masses from \citet{DeBreuck10}, 
the molecular gas fraction of HzRGs at $z\sim2$ is measured ranging from a few to 20\%, much lower than that of the star-forming main sequence with $M_{*}\sim10^{11}M_{\odot}$, at the same range of redshifts \citep[][for a review]{Scoville16,Tacconi18,Tacconi20}. Also, the molecular gas fraction of HzRGs at $z\sim$ 3--4 is $\sim$ 0.2--0.3, lower than that of the main sequence with $M_{*}\sim10^{11}M_{\odot}$ as well (Figure \ref{fig:fgvsz}).

In terms of the HzRGs with CO detections, \cite{Emonts14} have reported that in several CO(1--0) detected HzRGs, the CO line emission is spatially offset from their galaxy centers (off-center CO) and located along the radio jet axis (radio-CO alignment). They find marginal statistical significance that such radio-CO alignment is present in HzRGs. In addition to mergers, some causal connections between the jets and the gas reservoirs have also been suggested. For instance, jet-induced star formation, an important physical process in the positive feedback, appears to be related to these spatial properties \citep[e.g.,][]{Klamer04, Emonts14}.

TN J0924$-$2201 (hereafter TN J0924) is one of the most distant radio galaxies, at $z=5.1989$, determined by the measurement of Ly$\alpha$ \citep{vanBreugel99}.
\citet{Klamer05} have reported the detection of CO(1--0) and CO(5--4) line emission and the corresponding massive ($10^{11}$ $M_{\rm \odot}$) molecular gas reservoir, in TN J0924 by using the Australia Telescope Compact Array (ATCA). 
They find that CO(1--0) emission shows a tentative spatial offset of $\sim4''.5$ north to the nucleus of TN J0924, but the spatial offset is smaller than the beam size of $14''.5 \times 10''.1$. 
On the other hand, CO(5--4) emission shows a tentative spatial offset of $\sim2''.8$ south to the radio galaxy. The offset is also less than the beam size.
In addition,
combining with stellar mass of $10^{11.1}$ $M_{\rm \odot}$ estimated with mid-infrared (IR) data by \citet{DeBreuck10}, the molecular gas fraction of TN J0924 can be derived as 0.44, which is high in HzRGs (Figure \ref{fig:fgvsz}). 
As TN J0924 is the only CO-observed HzRG at $z\sim5$ and showing many intriguing properties mentioned above, we conducted further CO(1--0) observations of TN J0924 with the \textit{Karl G. Jansky} Very Large Array (VLA), to investigate the distribution of associated molecular gas and the reason why the molecular gas fraction of this object is apparently high.
We here present our new VLA observations of TN J0924 that provides the highest spatial resolution CO(1--0) imaging, reaching $\sim2''$. We also use archival Atacama Large Millimeter/submillimeter Array (ALMA) data in this work to constrain the rest-frame far-infrared properties of the system.
In Section~\ref{sec:obs}, we describe the observations and data.
The results are shown in Section~\ref{sec:results}. We discuss the results in Section~\ref{sec:discussion}, and conclude in Section~\ref{sec:conclusions}.
Throughout this paper, we 
assume a standard $\Lambda$CDM cosmology with $H_{0}=70\,\rm km\,s^{-1}\,Mpc^{-1}$, $\Omega_{\rm M}=0.3$ and $\Omega_{\rm \Lambda}=0.7$. 
At $z=5.2$, 1$''$ corresponds to $\sim$6.2 kpc.

\section{Observations and data} \label{sec:obs}

\subsection{VLA observations} \label{sec:vlaobs}

Our VLA K-band observations of TN J0924 were conducted in semester 20A (Project ID: 20A-303, PI: K. Lee), as a part of observations of seven HzRGs at $z>4.5$ (K. Lee et al. in preparation). The observations were conducted with 27 antennas in the array configuration C, with the baseline length ranging from 45 m to 3.4 km. 
The primary calibrator (bandpass and amplitude) is 3C138 and the secondary  calibrator (phase) is J0921--2618. 
Three executions for a 140-minute-long scheduling block had been carried out on 2020 February 16, 25 and March 07, resulting in an on-source integration time of 300 minutes. 

The K-band receiver was tuned to cover the frequency of the redshifted CO(1--0) line which is estimated from its redshift of Ly$\alpha$. We noted that the CO(1--0) line emission is at the redward of $z_{\rm Ly\alpha}$ with a velocity offset $\sim$150 $\rm km\,s^{-1}$\citep{Klamer05}. The total bandwidth was 2 GHz, covered from 18.09 to 20.11 GHz. It divided into 16 spectral windows with bandwidth of 128 MHz, and the spectral resolution is 2 MHz ($\sim$32 $\rm km\,s^{-1}$).
In VLA data, there are small spectral gaps between the spectral windows where the sensitivity drops. One of the 6-MHz-wide gaps between two spectral windows locates close to the targeted frequency of redshifted CO(1--0) (18.5954 GHz). Three channels (18.5977--18.6037 GHz) at the edge of spectral windows need to be flagged out. This leads to a blank spectral region with width $\sim$96 $\rm km\,s^{-1}$ in the spectrum close to $z_{\rm Ly\alpha}$.
Basic information of TN J0924 and descriptions of our VLA observations are summarized in Table \ref{tab:vlaobs}.

\begin{deluxetable}{lc}[ht!]
\tablenum{1}
\tablecaption{VLA observations\label{tab:vlaobs}}
\tablecolumns{2}
\tablewidth{0pt}
\tablehead{
\colhead{} & \colhead{TN J0924--2201}
}
\startdata
RA (J2000) & 09$^{\rm h}$24$^{\rm m}$19$^{\rm s}$.90 \\
Dec (J2000) & $-$22$^{\circ}$01$'$41$''$.4 \\
$z_{\rm Ly\alpha}$ & 5.1989 \\ 
\hline
Baseline (m) & 45--3400\\
On-source time (minute) & 300\\
Primary calibrator & 3C138\\
Secondary calibrator & J0921--2618\\
Frequency coverage (GHz) & 18.09--20.11\\
 & 2020-02-16\\
Date & 2020-02-25\\
 & 2020-03-07\\
\enddata
\tablecomments{This is the optical peak position  (identified from $HST$/ACS $i$-band image; Figure \ref{fig:hst+}).
}
\end{deluxetable}

Data reduction was performed with Common Astronomy Software Applications \citep[CASA;][]{CASA22} version 5.6.1 in the standard manner. All images were produced with the CASA task {\tt tclean}. The continuum map was produced with deconvolver {\tt mtmfs}, being cleaned down to $5\sigma$ level, and {\tt nterms} = 2, because the ratio of bandwidth to frequency is relatively large. Briggs weighting with {\tt robust} = 0.5 was applied. The synthesized beam size is 1$''$.8 $\times$ 0$''$.9 (Position angle P.A. = --6$^{\circ}$.7). The rms noise level is 4.3 $\rm \mu Jy\,beam^{-1}$.
To obtain the spectral cubes, we used CASA task {\tt uvcontsub} to subtract the continuum. All the channels were used for the fit, and we set {\tt fitorder} $= 1$.
We then applied {\tt tclean} on the continuum-subtracted measurement set with deconvolver {\tt hogbom}. We did not clean the cube because after the subtraction of continuum, there is no strong source in the data cube. Briggs weighting with {\tt robust} = 2.0 was applied.
We adopted the native spectral resolution of 2 MHz, which corresponds to the velocity resolution of $\sim$32 $\rm km\,s^{-1}$. Our spectral cubes were produced in two different synthesized beam sizes, one is 2$''$.3 $\times$ 1$''$.2 (P.A. = --7$^{\circ}$.0) and the other is 4$''$.4 $\times$ 2$''$.3 (P.A. = --7$^{\circ}$.0) which was smoothed by using {\tt imsmooth}, and the resulting rms noise level is 46.8 $\rm \mu Jy\,beam^{-1}\,channel^{-1}$ and 66.2 $\rm \mu Jy\,beam^{-1}\,channel^{-1}$, respectively, where one channel corresponds to $\sim32$ $\rm km\,s^{-1}$.
The pixel scale of all VLA images is set to $0''.3$.
Descriptions of images are summarized in Table \ref{tab:image}.

\begin{deluxetable}{lc}[ht!]
\tablenum{2}
\tablecaption{VLA and ALMA images\label{tab:image}}
\tablecolumns{1}
\tablewidth{0pt}
\tablehead{
\colhead{VLA continuum map}
}
\startdata 
Central frequency (GHz) & 19\\
Beam size & 1$''$.8 $\times$ 0$''$.9\\
Position angle & --6$^{\circ}$.7\\
rms noise ($\rm \mu Jy\,beam^{-1}$) & 4.3\\
\cutinhead{VLA spectral cube}
Beam size & 2$''$.3 $\times$ 1$''$.2\\
Position angle & --7$^{\circ}.0$\\
rms noise ($\rm \mu Jy\,beam^{-1}\,channel^{-1}$) & 46.8\\
\hline
Beam size & 4$''$.4 $\times$ 2$''$.3\\
Position angle & --7$^{\circ}.0$\\
rms noise ($\rm \mu Jy\,beam^{-1}\,channel^{-1}$) & 66.2\\
\cutinhead{ALMA continuum map}
Central frequency (GHz) & 230\\
Beam size & 1$''$.6 $\times$ 1$''$.0\\
Position angle & --81$^{\circ}.6$\\
rms noise ($\rm \mu Jy\,beam^{-1}$) & 58.9\\
\enddata
\tablecomments{one channel $\sim32$ $\rm km\,s^{-1}$.}
\end{deluxetable}

\subsection{ALMA data}

\begin{figure*}[ht!]
\epsscale{1.15}
\plotone{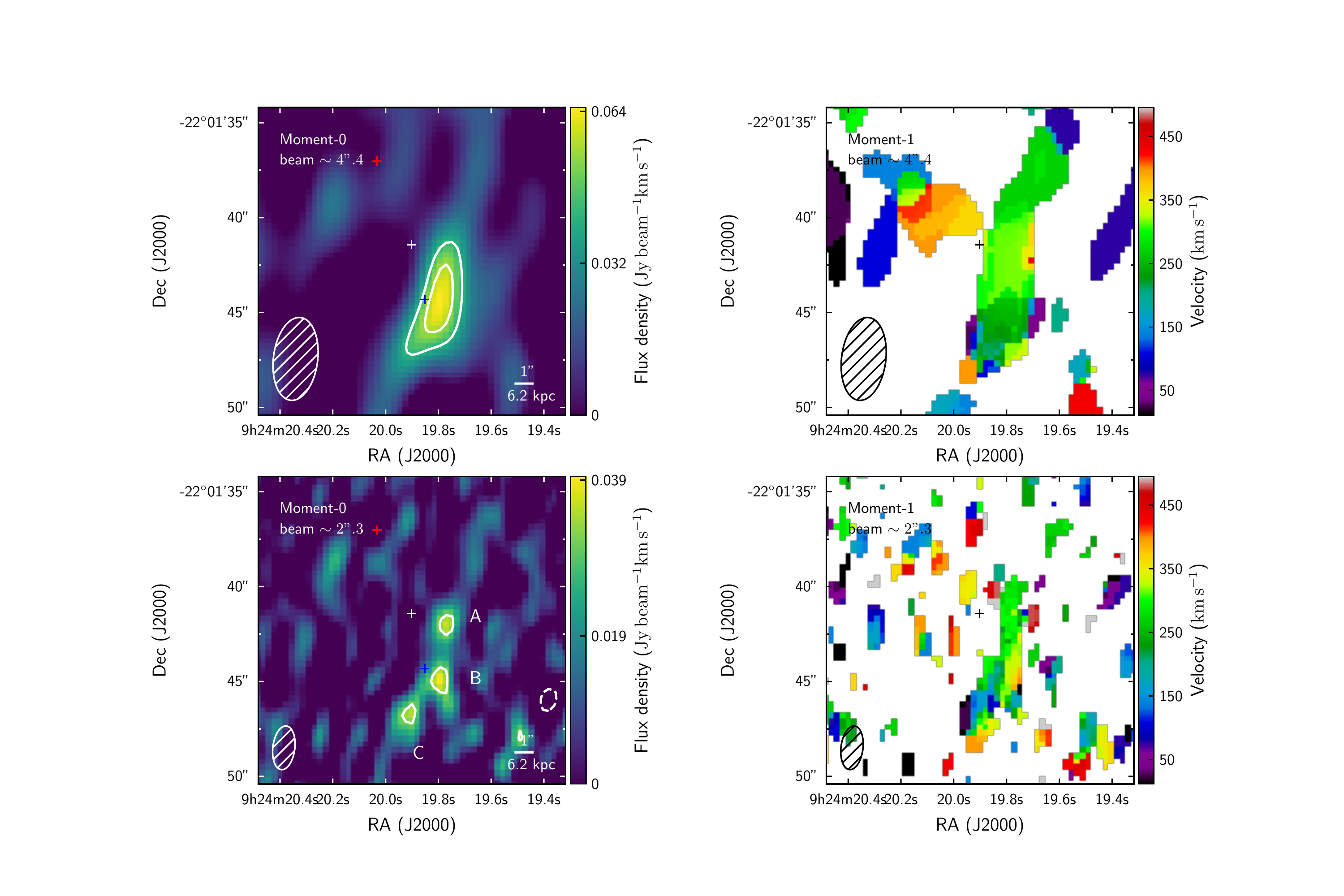}
\caption{The velocity-integrated (moment-0) CO(1--0) images and the flux-weighted velocity distributions (moment-1) of CO(1--0) line. The left two panels show the moment-0 images, and the white contours indicate $3\sigma$ and $4\sigma$, where in the top and the bottom image, $1\sigma=0.014$ $\rm Jy\,beam^{-1}\,km\,s^{-1}$ and $1\sigma=0.010$ $\rm Jy\,beam^{-1}\,km\,s^{-1}$, respectively.
The right two panels show the moment-1 images with $1\sigma$ cutoff. The integration range of velocity is 0--500 $\rm km\,s^{-1}$. The upper two panels show the images with beam size $\sim4''4$, and the bottom two panels show the images with beam size $\sim2''3$.
The white and the black crosses indicate the optical position, which is considered as the galaxy center of TN J0924. The red and the blue crosses in the left two panels indicate the peaks of CO(1--0) and CO(5--4), respectively, from \citet{Klamer05}. 
The ellipse at the bottom left corner in each image is the synthesized beam. In the bottom left image, white letter A, B and C indicate the clump A, B and C, respectively.
\label{fig:moment}}
\end{figure*}

The archival millimeter-wave data are from ALMA Band-6 observations (Project ID: 2013.1.00039.S). 
The on-source integration time is about 8 minutes.
The total bandwidth was 7.5 GHz, covered from 221.60 to 238.48 GHz ($\sim$230 GHz, corresponding to wavelength $\sim1.3$ mm). It divided into four spectral windows with a bandwidth of 1.875 GHz.
Since the data reduction needs to be ran with the ALMA pipeline that is included in the old CASA version 4.2 or 4.3, we requested to East Asian ALMA Regional Center for data reduction support. After receiving the calibrated measurement set, we used CASA version 6.2.0 for the further analysis.
We used CASA task {\tt tclean} with deconvolver {\tt hogbom}, cleaning down to $3\sigma$ level, to obtain the continuum image. Briggs weighting with {\tt robust} = 2.0 was applied. The synthesized beam size is $1''.6 \times 1''.0$ (P.A. = $-81^{\circ}.6$). The rms noise level is 43.1 $\rm \mu Jy\,beam^{-1}$.
The pixel scale is set to $0''.1$.
Descriptions of this ALMA continuum image are summarized in Table \ref{tab:image}.

\section{Results} \label{sec:results}

\subsection{CO(1--0) line and 19 GHz continuum}

Figure \ref{fig:moment} shows the velocity-integrated (moment-0) CO(1--0) images and the flux-weighted velocity distributions (moment-1) of CO(1--0) line.
We integrated the velocity from 0 to 500 $\rm km\,s^{-1}$ in the spectral cubes to produce the moment images by using CASA task {\tt immoments}. Two sets of images in different angular resolutions of $4''.4$ and $2''.3$ are shown. 
The images with a larger beam size of $4''.4$ are shown as the comparison with the ATCA detection reported by \citet{Klamer05}. 
We used two-dimensional (2D) Gaussian fit to estimate the CO(1$-$0) line velocity-integrated intensity. The CO(1$-$0) line velocity-integrated intensity from our $4''.4$ image is $0.078\pm0.016$ $\rm Jy\,km\,s^{-1}$, which is
comparable to their ATCA result, $0.087\pm0.017$ $\rm Jy\,km\,s^{-1}$, and the resulting molecular gas mass is comparable as well.
The peak of the CO(1--0) emission at 4".4 resolution is $\sim4''$ offsetting from the optical position of TN J0924.
\citet{Klamer05} have reported that CO(1--0) emission appears to be tentatively $4''.5$ offset to the north from the center of the host galaxy with $\sim 10''$ beam. 
Our results confirm that the spatial offsets between CO(1--0) and the nucleus of the galaxy exist, while the spatial offsets are generally in the southwest. However, these are similar with their result of CO(5--4), which is $\sim2''.8$ offset to the south. It is reasonable since the beam size of their CO(5--4) observation is about $5''$, smaller than their $10''$ beam of CO(1--0) observation.

From the moment-0 images with $4''.4$ beam, the CO(1--0) emission is elongated in north-south direction, similar to the shape of the beam but appears to be partially resolved. While in the image with a $2''.3$ beam, the CO(1--0) emission is resolved into three clumps. We label them as clump A, B and C from north to south. 

Given the moment-1 (with 1$\sigma$ cutoff) images, which is suffered from low signal-to-noise (S/N) ratios, it is difficult to tell the motions of three clumps. In the image with either $4''.4$ beam or $2''.3$ beam, it appears that the velocity distribution is centered around 300 $\rm km\,s^{-1}$ with little dispersion. 
Nevertheless, we need deeper CO observations to study the velocity structure of the system.  

\begin{figure}[ht!]
\epsscale{1.25}
\plotone{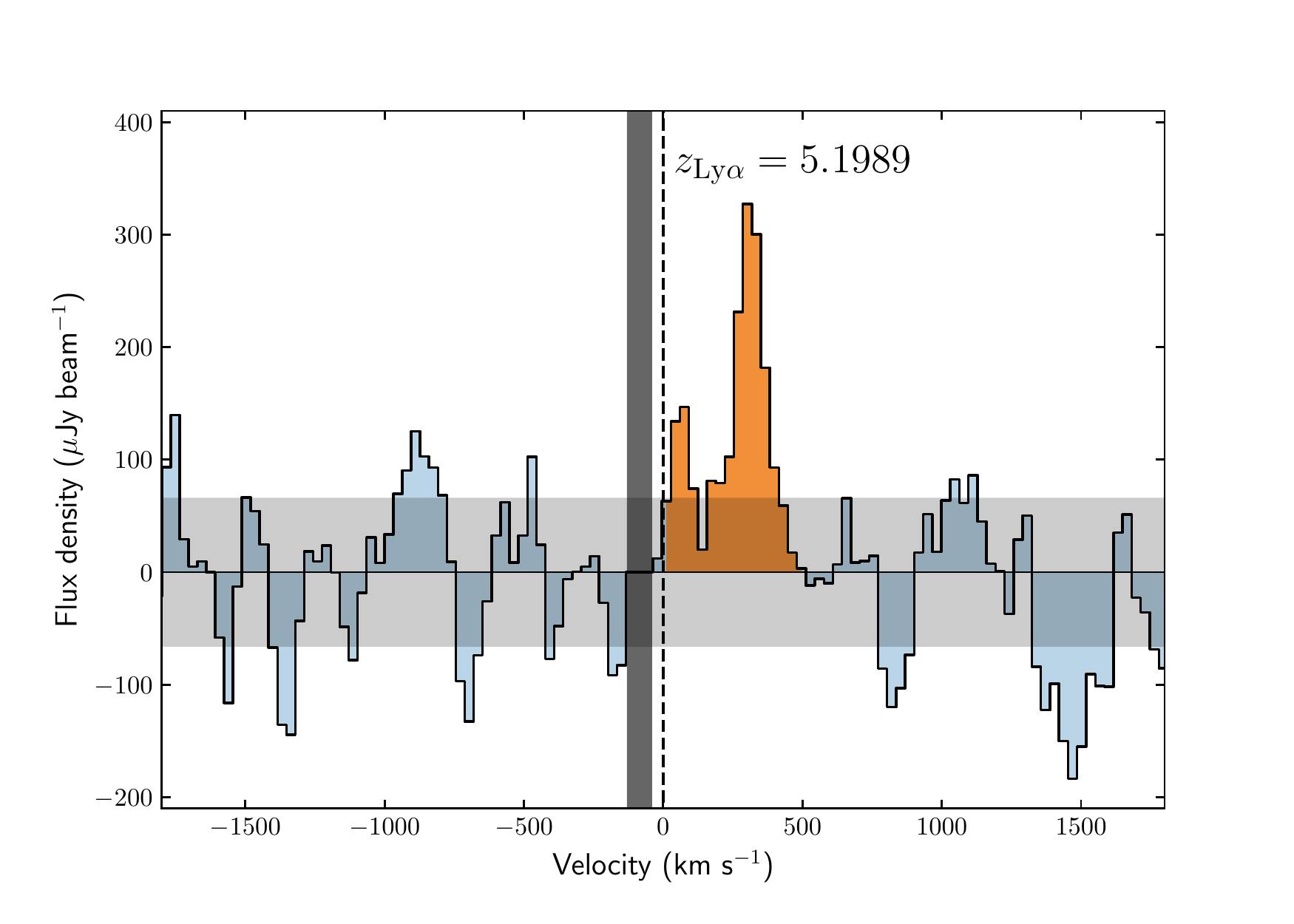}
\caption{The CO(1--0) spectrum of TN J0924$-$2201 in the spectral cube whose beam size is $4''.4$. Dashed line indicates the redshift $z_{\rm Ly\alpha}=5.1989$ derived from Ly$\alpha$ \citep{vanBreugel99}. The orange area shows the velocity range we integrated to produce the velocity-integrated map.
The shallow shaded area indicates the $1\sigma$ noise level and the dark shaded area represents the blank region in the data. The spectrum is extracted from the peak of the emission.
\label{fig:spectrum}}
\end{figure}

Figure \ref{fig:spectrum} shows the detected CO(1--0) line in the spectral cube whose beam size is $4''.4$. The S/N ratio of the peak is 4.6. The dark shaded area in Figure \ref{fig:spectrum} is the data blank in our spectrum due to the sensitivity drops between the spectral windows which is mentioned in Sec \ref{sec:vlaobs}. Nonetheless, since the detection appears redward of $z_{\rm Ly\alpha}$ and the flux close to the blank channels are negative, we consider the blank region does not affect on the significance of CO(1--0) detection in our data.

As \citet{Klamer05} reported, the detection of CO(1--0) is redward of $z_{\rm Ly\alpha}$, and there are two kinematic features apparent in the spectrum. However, the brighter and dominant component we find is centering at $v=300$ $\rm km\,s^{-1}$, which is similar with the weaker feature which \citet{Klamer05} reported, but $\sim$150 $\rm km\,s^{-1}$ redder than the dominant feature they found. On the other hand, the weaker feature in our spectrum centered at $v=100$ $\rm km\,s^{-1}$, where is $\sim$50 $\rm km\,s^{-1}$ bluer than their dominant feature. We treated both of features as the detection, and integrated them over 0 to 500 $\rm km\,s^{-1}$ to produce the velocity-integrated map. 
For the comparison, we also smoothed our VLA data to the ATCA beam size of $14''.5\times10''.1$, and find that the peak flux density of signal is $\sim$0.5 mJy$\,$beam$^{-1}$, which is consistent with the ATCA result in \citet{Klamer05}. But the stronger line feature is still on the red side and the same spatial location as what we analyzed with our VLA data cubes in higher angular resolutions. However, the S/N only reaches $\sim$2 ($1\sigma$ noise $\sim$0.25 mJy$\,$beam$^{-1}$) when we smoothed the data to $14''.5\times10''.1$.

\begin{figure}[ht!]
\epsscale{1.25}
\plotone{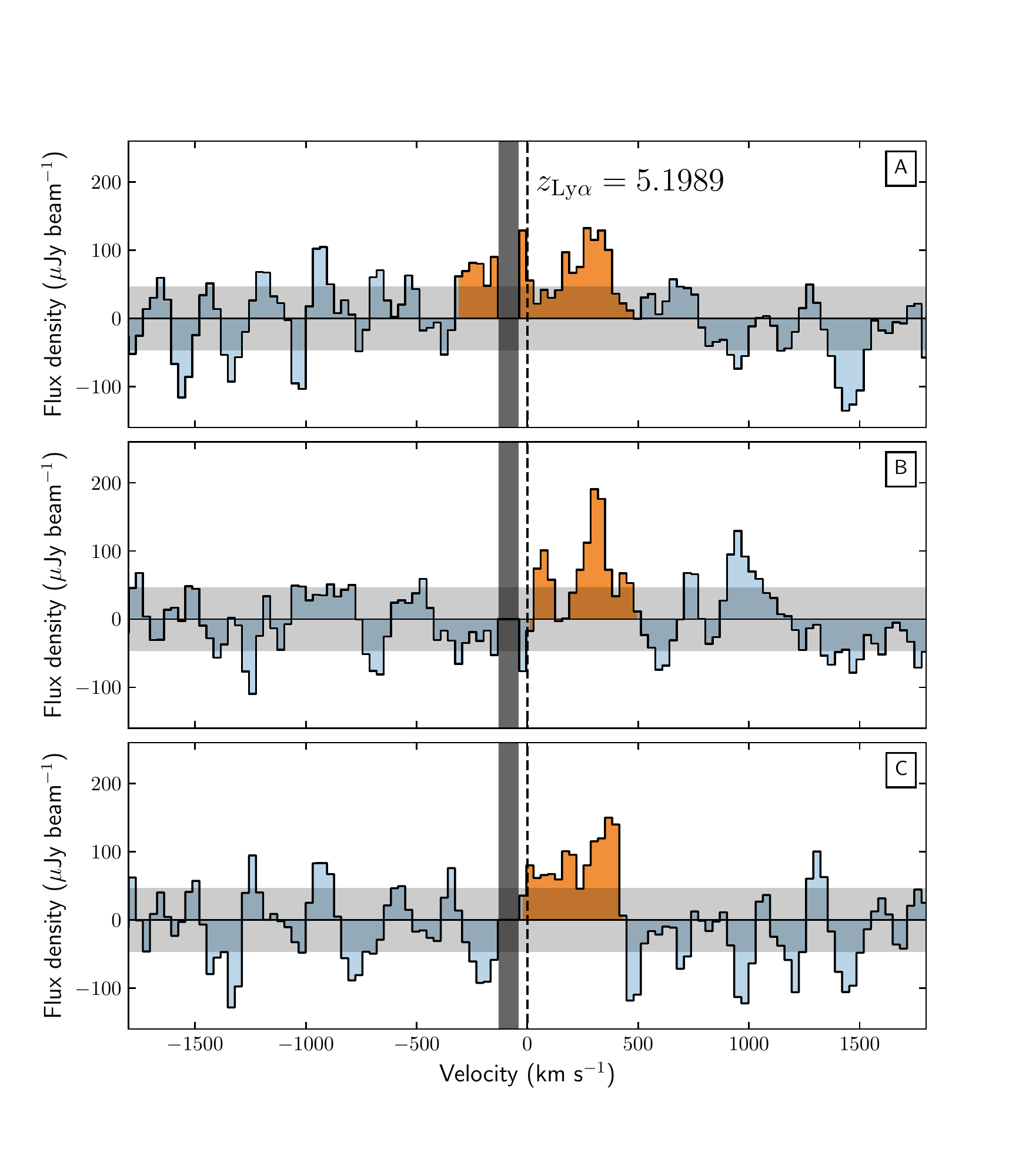}
\caption{The CO(1--0) spectra of TN J0924$-$2201 in the spectral cube whose beam size is $2''.3$. The upper, middle and lower panels show the spectra of clump A, B and C, respectively. Dashed line indicates the redshift $z_{\rm Ly\alpha}=5.1989$ derived from Ly$\alpha$ \citep{vanBreugel99}. The orange areas show the velocity range we integrated to produce the velocity-integrated map.
The shallow shaded areas indicate the $1\sigma$ noise level and the dark shaded areas represent the blank region in the data. The spectra were extracted from the peak of the emission in each region.}
\label{fig:spectrumABC}
\end{figure}

Figure \ref{fig:spectrumABC} shows the spectra of clump A, B and C. The CO(1--0) line in each clump generally peaks at $v=300$ $\rm km\,s^{-1}$, consistent with the velocity of the dominant component at the angular resolution of $4''.4$ (Figure $\ref{fig:spectrum}$). We found that in the clump A, while there is a blank region in the spectrum, it appears that the detection of CO(1--0) is broader than those in clump B and C. To get the highest S/N of each clumps on the velocity-integrated maps, different ranges of velocity for integration was applied, since there are some velocity differences between three clumps. 
In clump A, the flux within the data blank is integrated as zero.

\begin{figure*}[ht!]
\epsscale{1.0}
\plotone{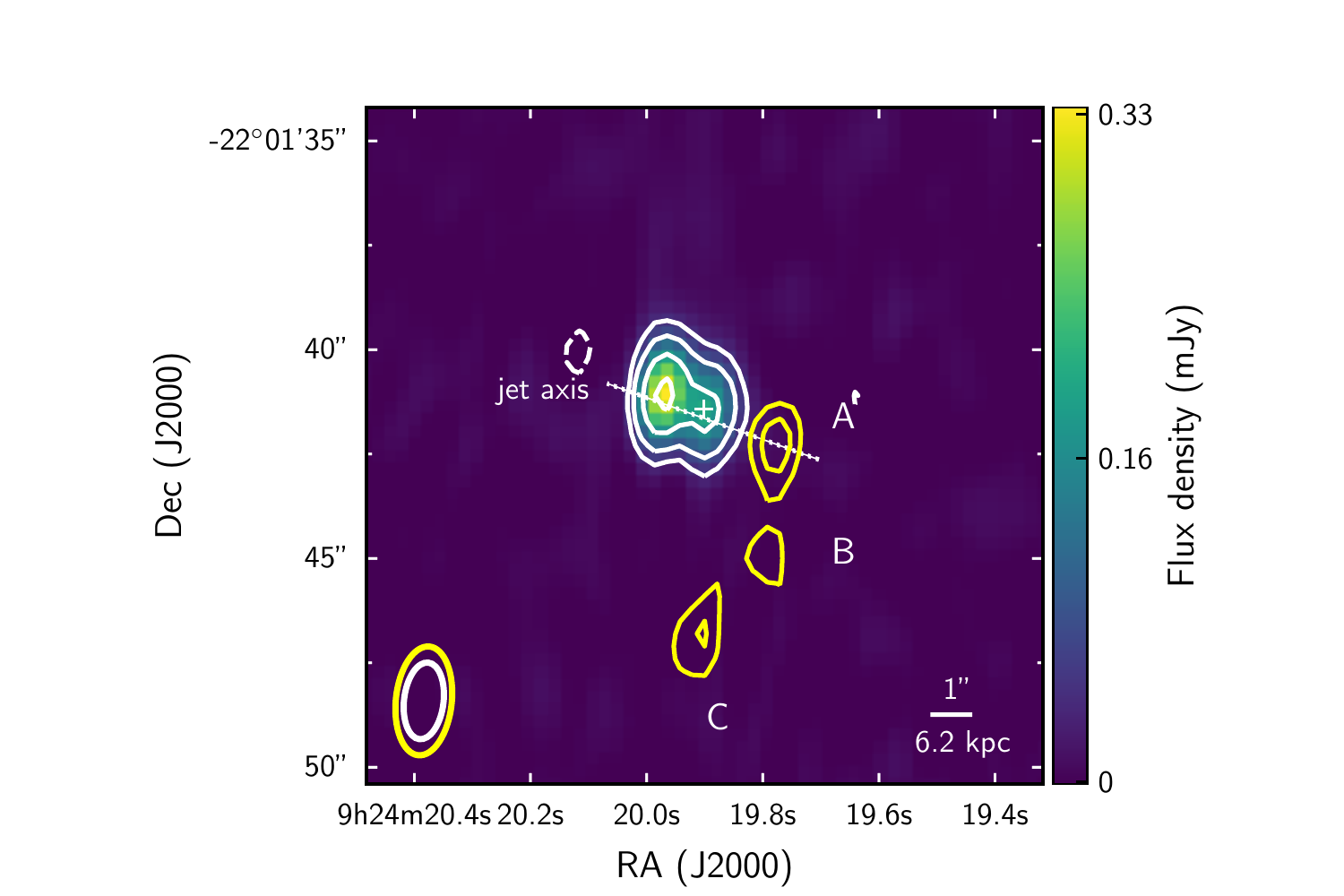}
\caption{Colored 19 GHz continuum map with CO(1--0) velocity-integrated emission in $2''.3$ beam overlaid in yellow. Yellow contours indicate 3$\sigma$ and 4$\sigma$, where for clump A, B, and C, $1\sigma=0.013$ $\rm Jy\,beam^{-1}\,km\,s^{-1}$, $1\sigma=0.010$ $\rm Jy\,beam^{-1}\,km\,s^{-1}$ and $1\sigma=0.013$ $\rm Jy\,beam^{-1}\,km\,s^{-1}$, respectively. White contours indicate $10\sigma$, $20\sigma$, $40\sigma$ and $70\sigma$, where $1\sigma=4.3$ $\mu$Jy. Dashed contours show -3$\sigma$. The white cross indicates the optical position. The white dotted line indicates the jet axis.
The ellipses at the bottom-left corner are the synthesized beams, which correspond to the continuum map and CO(1--0) velocity-integrated emission in white and yellow, respectively. White letter A, B and C indicate the clump A, B and C, respectively. The integration velocity range of each clump is shown in Figure \ref{fig:spectrumABC} in orange.
\label{fig:cont_mom0}}
\end{figure*}

Figure \ref{fig:cont_mom0} shows spatial distributions of 19 GHz continuum image and the CO(1--0) velocity-integrated emission in $2''.3$ beam. There are apparent spatial offsets between the continuum emission and CO(1--0) line emissions. 

\begin{deluxetable*}{lccc}[ht!]
\tablenum{3}
\tablecaption{Physical properties of observed CO(1--0) line emission \label{tab:COline}}
\tablecolumns{4}
\tablewidth{0pt}
\tablehead{
\colhead{} & \colhead{Clump A} & \colhead{Clump B} & \colhead{Clump C}
}
\startdata
RA (J2000)  & 09$^{\rm h}$24$^{\rm m}$19$^{\rm s}$.77 & 09$^{\rm h}$24$^{\rm m}$19$^{\rm s}$.79 & 09$^{\rm h}$24$^{\rm m}$19$^{\rm s}$.90\\
Dec (J2000)  & $-$22$^{\circ}$01$'$42$''$.3 & $-$22$^{\circ}$01$'$45$''$.0& $-$22$^{\circ}$01$'$46$''$.8\\
$\Delta$RA$^{a}$ (arcsec)  & $-1.8$ & $-1.5$ & $+0.0$\\
$\Delta$Dec$^{b}$ (arcsec)  & $-0.9$ & $-3.6$ & $-5.4$\\
Peak ($\rm \mu Jy\,beam^{-1}$) & $132.3\pm46.8$ & $190.6\pm46.8$ & $149.9\pm46.8$\\
Velocity width$^{c}$ ($\rm km\,s^{-1}$)  & 840 & 500 & 450 \\
$I_{\rm CO(1-0)}$$^{d}$ ($\rm Jy\,km\,s^{-1}$)  & $0.050\pm0.008$ & $0.034\pm0.006$ & $0.039\pm0.006$ \\
$L'_{\rm CO(1-0)}$ ($\rm K\,km\,s^{-1}pc^{2}$)  & $(4.7\pm0.7)\times10^{10}$ & $(3.2\pm0.5)\times10^{10}$ & $(3.7\pm0.5)\times10^{10}$ \\
$M_{\rm H_{2}}$$^{e}$ ($M_{\odot}$) & $
(3.7\pm0.6)\times10^{10}$ & $(2.5\pm0.4)\times10^{10}$ & $(3.0\pm0.4)\times10^{10}$\\
\enddata
\centering
\tablecomments{
$^{a}$Relative to RA: 09$^{\rm h}$24$^{\rm m}$19$^{\rm s}$.90, where is the optical peak.\\
$^{b}$Relative to Dec: $-$22$^{\circ}$01$'$41$''$.4, where is the optical peak.\\
$^{c}$Referring to Full Width at Zero Intensity (FWZI).\\
$^{d}$The uncertainty is from the 2D Gaussian fit.\\
$^{e}$Corresponding to $\alpha_{\rm CO} = 0.8$ $M_{\rm \odot}\rm\,(K\,km\,s^{-1}pc^{2})^{-1}$.
}
\end{deluxetable*}

The 19 GHz continuum is considered to be dominated by the non-thermal synchrotron emission from the radio jet. It is partially resolved into two components, east and west. 
We used 2D Gaussian fit to estimate the integrated flux on each side.
The peak and the integrated flux density of the eastern component is $0.333\pm0.004$ mJy and $0.451\pm0.030$ mJy, respectively, brighter than that of the western component, whose peak and the integrated flux density is $0.206\pm0.004$ mJy and $0.347\pm0.015$ mJy, respectively.
The uncertainty is from the 2D Gaussian fit. We also smoothed the continuum image to $14''.5 \times 10''.1$, and the integrated flux density is $0.715\pm0.038$ mJy, well consistent with the measurement reported by \cite{Klamer05}.
The radio jet axis is in east-west direction, almost aligning with the clump A.
Therefore, besides the confirmation of off-center CO, we observed the radio-CO alignment in TN J0924 for the first time.

The velocity-integrated flux density of CO(1--0) emission $I_{\rm CO(1-0)}$ at clump A, B and C are $0.050\pm0.008$ $\rm Jy\,km\,s^{-1}$, $0.034\pm0.006$ $\rm Jy\,km\,s^{-1}$ and $0.039\pm0.006$ $\rm Jy\,km\,s^{-1}$, respectively.
The uncertainty is from the 2D Gaussian fit.
And the corresponding CO line luminosity $L'_{\rm CO(1-0)}$ of clump A, B and C are $(4.7\pm0.7) \times10^{10}$ $\rm \,K\,km\,s^{-1}pc^{2}$, $(3.2\pm0.5) \times10^{10}$ $\rm \,K\,km\,s^{-1}pc^{2}$ and $(3.7\pm0.5) \times10^{10}$ $\rm \,K\,km\,s^{-1}pc^{2}$, respectively. 
These physical properties and the spatial offsets between three clumps and the nucleus, which is assumed to locate at the optical peak (RA: 09$^{\rm h}$24$^{\rm m}$19$^{\rm s}$.90, Dec: $-$22$^{\circ}$01$'$41$''$.4), are summarized in Table \ref{tab:COline}.
On the other hand, with the assumption that the Full Width at Zero Intensity (FWZI) of line emission is 500 $\rm km\,s^{-1}$, the $3\sigma$ upper limit for non-detection is $1.7\times10^{10}$ $\rm \,K\,km\,s^{-1}pc^{2}$.

\subsection{ALMA 1.3 mm continuum}

\begin{figure}[ht!]
\epsscale{1.3}
\plotone{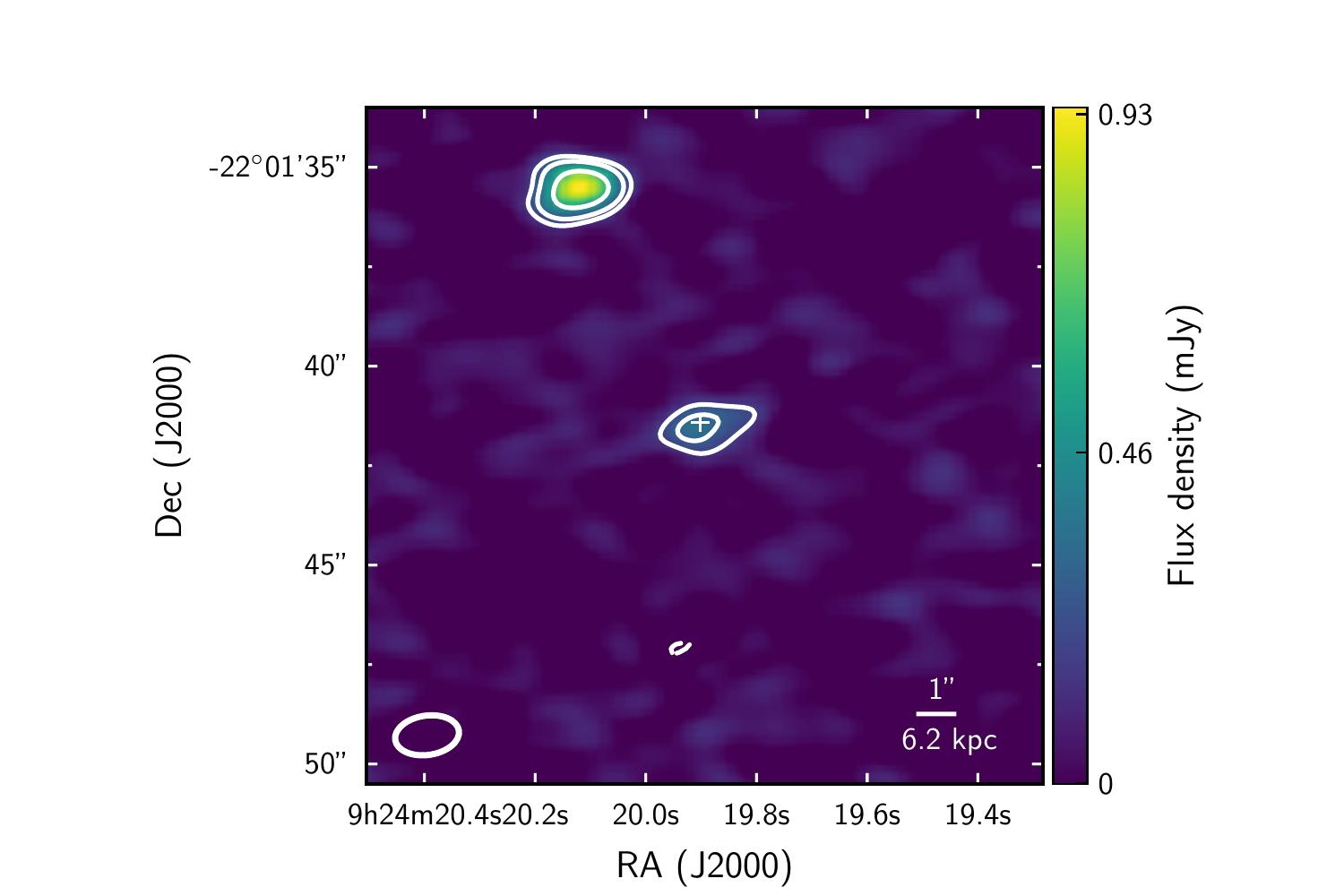}
\caption{Colored ALMA Band-6 continuum image. White contours indicate 3$\sigma$, 5$\sigma$ and 10$\sigma$, where $1\sigma=58.9$ $\mu$Jy. Dashed contours show -3$\sigma$. The white cross indicates the optical position. The ellipse at the bottom-left corner is the synthesized beam.
\label{fig:alma_cont}}
\end{figure}

Figure $\ref{fig:alma_cont}$ shows the ALMA Band-6 (230 GHz; 1.3 mm) continuum image. The source at the center is considered to be the thermal dust emission from the host galaxy of TN J0924. The peak flux is $0.35\pm0.059$ mJy$\,\rm beam^{-1}$ and the integrated flux is $0.64\pm0.12$ mJy.

A source locates at the north of TN J0924 with a peak flux of $0.94\pm0.059$ mJy$\,\rm beam^{-1}$ and the integrated flux of $1.12\pm0.090$ mJy. 
At its position, no line emission is obtained in our VLA data, and no Ly$\alpha$ detection in \citet{Venemans04}.
It has a mid-infrared counterpart in IRAC 3.6 $\mu$m image, showing an integrated flux of $37.3\pm0.4$ $\mu$Jy. The measured 3.6 $\mu$m to 1.3 mm flux ratio, $3.3 \times 10^{-2}$, seem to suggest a redshift lower than TN J0924 \citep{Yamaguchi19}. Further constraints are needed to see whether the northern 1.3 mm continuum source is a part of the overdensity in the TN J0924 field at $z=5.2$. 

\subsection{Spectral energy distribution} \label{sec:SED}

\begin{figure}[ht!]
\epsscale{1.25}
\plotone{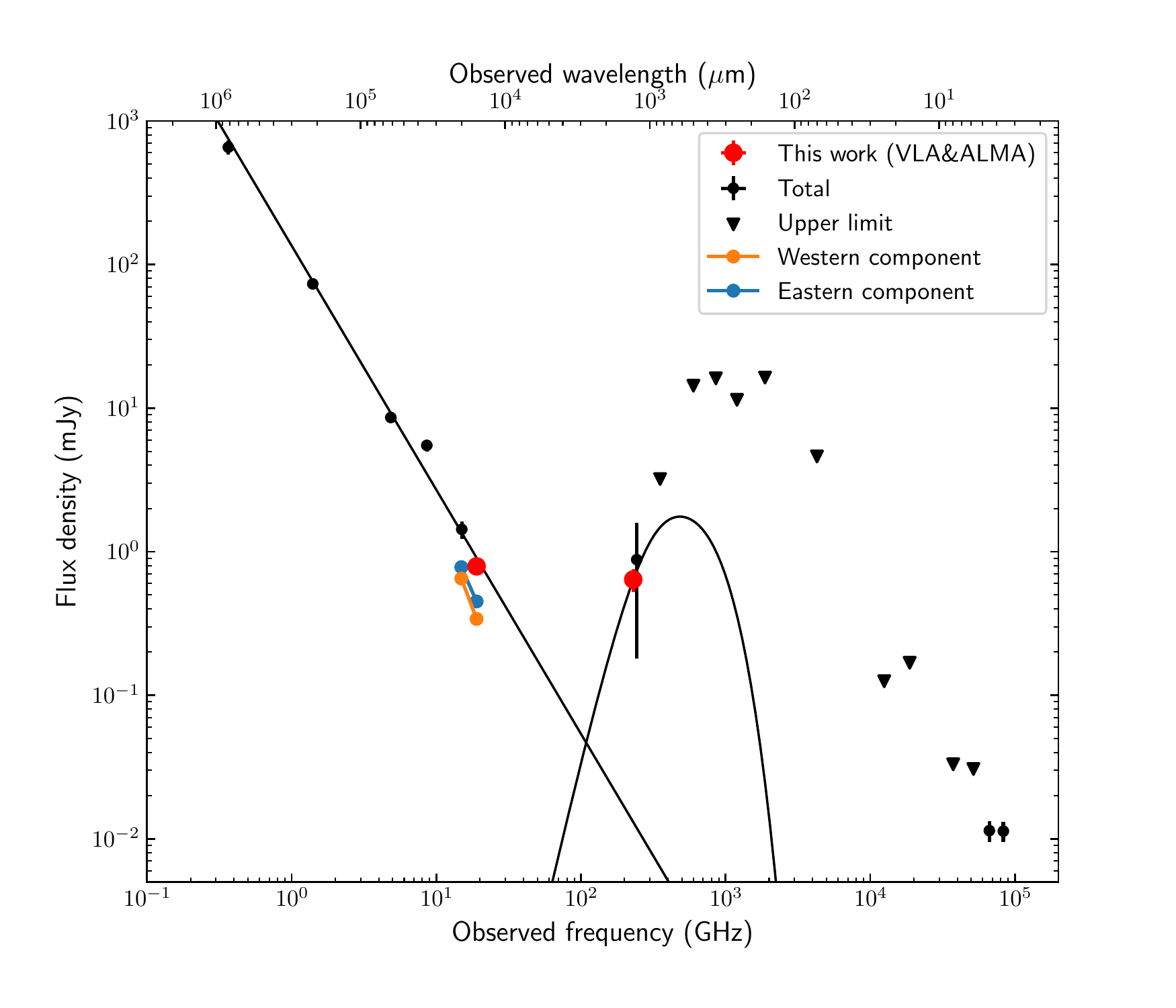}
\caption{Spectral energy distribution of TN J0924. Black dots and downward triangles represent total flux densities and $3\sigma$ upper limits, respectively, from IRAC, IRS, MIPS \citep{DeBreuck10}, PACS, SPIRE \citep{Drouart14}, SCUBA \citep{Reuland04}, WSRT, VLA \citep{DeBreuck00,Falkendal19} and ALMA \citep{Falkendal19}, in the literature.
Data in this work are plotted in red dots. The western and eastern components of radio continuum are indicated in orange and blue dots, respectively. The red dot represented VLA is simply the summation of the western and eastern components.
The power law of non-thermal synchrotron emission ($\alpha=-1.7$) and the normalized modified blackbody ($\beta=2.5$, $T_{\rm dust}=50$ K) are indicated in black solid lines.
\label{fig:sed}}
\end{figure}

Figure \ref{fig:sed} shows the spectral energy distribution (SED) of TN J0924. Observational data points are from IRAC, IRS, MIPS \citep{DeBreuck10}, PACS, SPIRE \citep{Drouart14}, SCUBA \citep{Reuland04}, WSRT, VLA \citep{DeBreuck00,Falkendal19}, ALMA \citep{Falkendal19} and this work (VLA \& ALMA).
From the extrapolation of the power law with spectral index $\alpha=-1.7$ (assuming $S_{\nu}\propto\nu^{\alpha}$),
which is derived from fluxes at 365 MHz and 19 GHz, 
non-thermal synchrotron emission is estimated to be about 2\% of the flux density of $0.64\pm0.117$ mJy observed at ALMA Band-6 (230 GHz). Therefore, we can confidently consider that the observed flux at 230 GHz is dominated by the thermal dust emission.

\begin{figure*}[ht!]
\epsscale{1.25}
\plotone{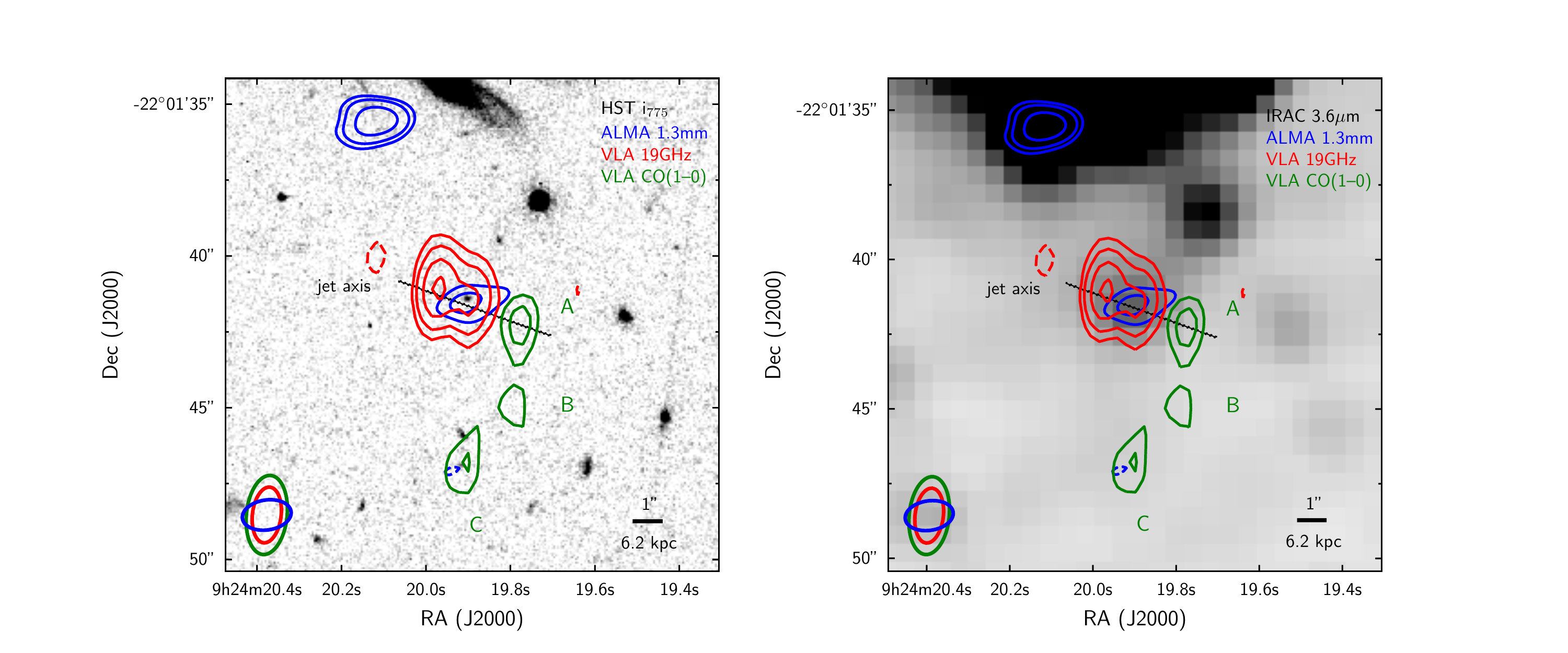}
\caption{The grey-scale $HST$/ACS F775W ($i$-band) map (left) and $Spitzer$/IRAC 3.6 $\mu$m map (right), with CO(1--0) velocity-integrated emission in $2''.3$ beam overlaid in green contours, 19 GHz continuum overlaid in red contours and ALMA  1.3 mm continuum overlaid in blue contours. Green contours indicate 3$\sigma$ and 4$\sigma$, where for clump A, B, and C, $1\sigma=0.013$ $\rm Jy\,beam^{-1}\,km\,s^{-1}$, $1\sigma=0.010$ $\rm Jy\,beam^{-1}\,km\,s^{-1}$ and $1\sigma=0.013$ $\rm Jy\,beam^{-1}\,km\,s^{-1}$, respectively. Red contours indicate 10$\sigma$, 20$\sigma$, 40$\sigma$ and 70$\sigma$, where $1\sigma=4.3$ $\mu$Jy. Blue contours indicate 3$\sigma$, 5$\sigma$ and 10$\sigma$, where $1\sigma=58.9$ $\mu$Jy. Dashed contours show -3$\sigma$. The black dotted line indicates the jet axis. The ellipses at the bottom-left corner are the synthesized beams corresponding to the contours in the same colors. 
\label{fig:hst+}}
\end{figure*}

Noticing that the ALMA data used by \citet{Falkendal19} is observed at 243 GHz with integration time of about 2 minutes,
the low S/N of detection from that data shows a large uncertainty of flux. Nevertheless, our robust ALMA detection at 230 GHz is consistent with their modified blackbody fitting.
By assuming a modified blackbody with the same
parameters they used --- emissivity spectral index $\beta=2.5$ and dust temperature $T_{\rm dust}=50$ K, 
we fit with our ALMA 230 GHz data and plot in Figure \ref{fig:sed}.

\section{Discussion} \label{sec:discussion}

\subsection{Spatial offsets}

Figure \ref{fig:hst+} shows the grey-scale 
$HST$/ACS F775W ($i$-band) image and $Spitzer$/IRAC 3.6 $\mu$m image overlaid with VLA 19 GHz continuum, CO(1--0) and ALMA 1.3 mm continuum contours. 
By comparing the position of a nearby star in Gaia catalog \citep{Gaia21}, 
we find that $HST$ image has little astrometric offset (RA: $-0''.17$, Dec: $+0''.06$). It is negligible for comparing $HST$ image with ALMA and VLA images.
It appears that at the center, the source on the $HST$ and IRAC images overlaps with the ALMA continuum and the western component of VLA 19 GHz continuum, indicating that it is the location of the host galaxy and the active galactic nucleus (AGN) of TN J0924.
We find that the clump A is about $2''.0$ (12 kpc) offsetting to the west of the nucleus, and apparently being aligning with the radio jet axis, which is roughly in east-west direction. Furthermore, the clump B and C are about $3''.9$ (24 kpc) and $5''.4$ (33 kpc) offsetting to the south west and the south, respectively. 
Spatial offsets between the molecular gas reservoir and the nuclei of HzRGs, and the alignment between molecular gas and the radio jet have been reported in lower-$z$ HzRGs studies \citep{Emonts14, Gullberg16a, Falkendal21}. 
Our results reveal that such spatial offsets and alignment between the molecular gas and radio jet happens in a $z>5$ environment. 

In TN J0924, the spatial distribution of the nucleus and three molecular gas clumps we find are special. All three clumps are off-center, but only clump A is on the jet axis. In addition, since the clump B and C are about 24 kpc and 33 kpc away from the nucleus, respectively, we suggest that only the clump A is a part of the radio galaxy, and the clump B and C are companion objects with massive molecular gas. 
In Figure \ref{fig:spectrumABC}, although the S/N of clump A is not high, the broad CO(1--0) line emission might support our suggestion. On the other hand, we do not find counterparts of clump A, B and C in $HST$, ALMA and IRAC images.

\subsection{Molecular gas and star formation rate}

\begin{figure*}[ht!]
\epsscale{1.2}
\plotone{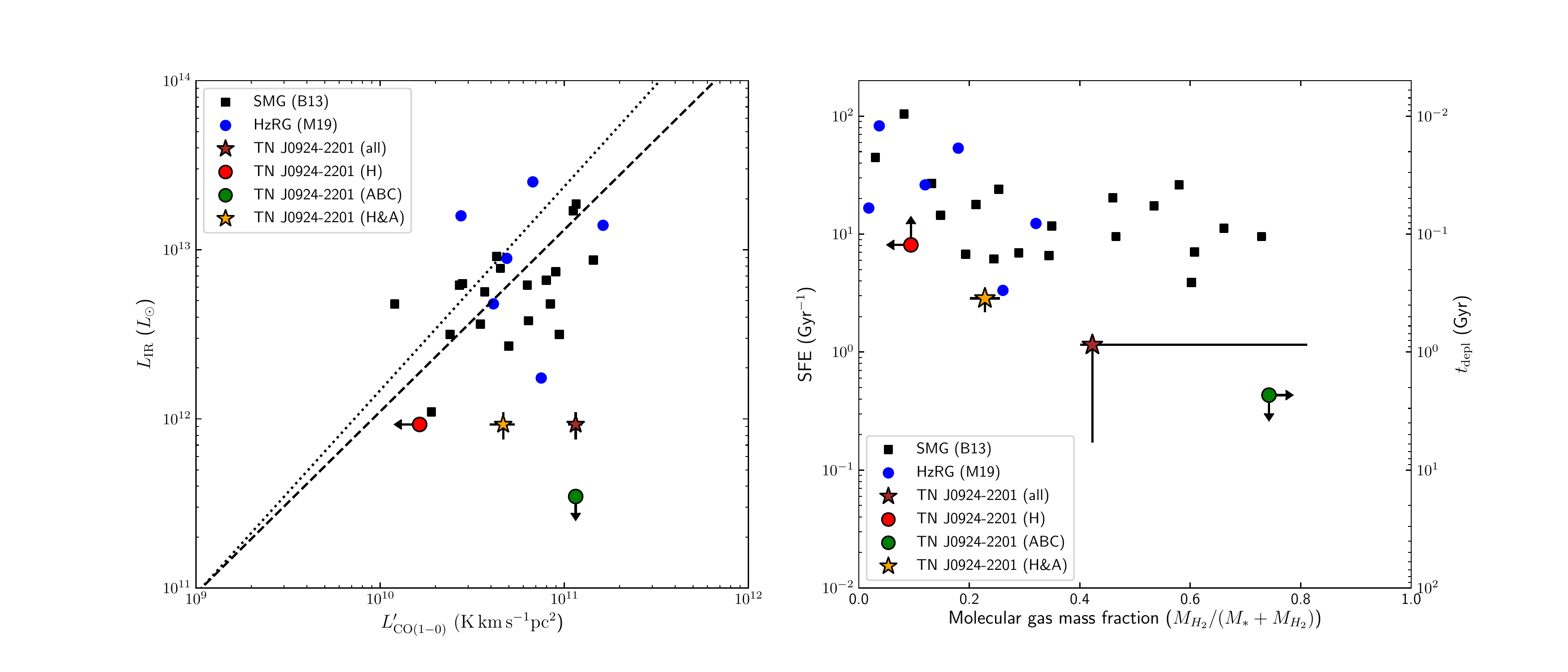}
\caption{Comparison of physical properties of TN J0924 with other SMGs and HzRGs. (a) The left panel shows $L_{\rm IR}$ versus $L'_{\rm CO(1-0)}$ (b) The right panel shows SFE and $t_{\rm depl}$ versus molecular gas fraction $f_{g}\equiv M_{H_{2}}/(M_{*}+M_{H_{2}})$.
In both panels, TN J0924 is plotted with four sets: TN J0924 (all) in brown star symbol indicates the host galaxy and three clumps; TN J0924 (H) in red dot indicates the host galaxy; TN J0924 (ABC) in green dot indicates three clumps; TN J0924 (H\&A) in orange star symbol indicates the host galaxy and the clump A. HzRGs at $z=2.1-3.8$ \citep[M19;][]{Man19} and SMGs at $z=1.4-4.1$ \citep[B13;][]{Bothwell13, Ivison11} are shown in blue dots and black squares, respectively. The black dashed line and dotted line indicate the fit of SMGs and the fit of SMGs together with (U)LIRGs, respectively, from \citet{Bothwell13}.
\label{fig:SFE_fg}}
\end{figure*}

In the literature, the CO-to-H$_{2}$ conversion factor $\alpha_{\rm CO}$ of ultra luminous infrared galaxies (ULIRG) is adopted for HzRGs. 
For the consistency with the previously reported molecular gas mass measurements of HzRGs including \citet{Klamer05}, 
we adopt $\alpha_{\rm CO} = 0.8$ $M_{\rm \odot}\rm\,(K\,km\,s^{-1}pc^{2})^{-1}$ \citep{DownesandSolomon98} 
to derive the molecular gas masses of three clumps. The molecular gas mass of clump A, B and C is $(3.7\pm0.6)\times10^{10}$ $M_{\rm \odot}$, $(2.5\pm0.4)\times10^{10}$ $M_{\rm \odot}$, and $(3.0\pm0.4)\times10^{10}$ $M_{\rm \odot}$, respectively. 
In total, this is about 90\% of the molecular gas mass derived by \citet{Klamer05}.
At the host galaxy, we constrain on the molecular gas mass with the $3\sigma$ upper limit of $1.3\times10^{10}$ $M_{\rm \odot}$.

We estimate the IR luminosity $L_{\rm IR}$ (rest-frame 8--1000 $\mu$m) and the star formation rate (SFR) of TN J0924 by following the assumption in \citet{Falkendal19}. 
They use modified blackbody originated from star formation to fit the monochromatic ALMA Band-6 (243 GHz) data. As mentioned in Section \ref{sec:SED}, in spite of the large uncertainty of $L_{\rm IR}$ and SFR in \citet{Falkendal19} due to their shallow ALMA data, the integrated flux of our robust ALMA Band-6 (230 GHz) detection (peak S/N $=8$) is consistent with their fitting. 
Therefore, we use the modified blackbody with the same physical parameters to fit our data and estimate $L_{\rm IR}$ to be $(9.3\pm1.7)\times10^{11}$ $L_{\rm \odot}$, which is consistent with their result, $1\times10^{12}$ $L_{\rm \odot}$. 
Compared to their data, the uncertainty of our ALMA data is smaller. And the result is reliable on the same order of magnitude. Changing $\beta$ and $T_{\rm dust}$ vary the $L_{\rm IR}$ by a factor of a few.
For instance, using $\beta=2.5$ and $T_{\rm dust}=40$ K makes $L_{\rm IR}=(3.6\pm0.7)\times10^{11}$ $L_{\rm \odot}$; using $\beta=1.5$ and $T_{\rm dust}=50$ K makes $L_{\rm IR}=(9.1\pm1.7)\times10^{11}$ $L_{\rm \odot}$.
Also, for consistency with the stellar mass derived by \citet{DeBreuck10}, we adopt a Kroupa IMF, resulting in the SFR of $110\pm20$ $M_{\rm \odot}\,\rm yr^{-1}$.
From the ALMA image, it appears that the dusty star formation activity highly associates with the host galaxy rather than with the molecular gas reservoirs. 
The dusty star formation activity appears to be confined to the host galaxy.
Thus, at the positions of three molecular gas clumps, we constrain on each of their $L_{\rm IR}$ and
SFR with the $3\sigma$ upper limit of $3.5\times10^{11} L_{\rm \odot}$ and 40 $M_{\rm \odot}\,\rm yr^{-1}$, respectively. 
The low upper limit of SFR leads to the concerns about the applied CO-to-H$_{2}$ conversion factor. We discuss this with Figure \ref{fig:SFE_fg} later.

In Figure \ref{fig:SFE_fg}, to show the properties of TN J0924 , we divide TN J0924 into four sets: (1) TN J0924 (all) indicates the host galaxy and three clumps as a whole; (2) TN J0924 (H) indicates only the host galaxy; (3) TN J0924 (ABC) indicates three clumps; (4) TN J0924 (H\&A) indicates the host galaxy and the clump A. HzRGs at $z=2.1-3.8$ \citep{Man19} and submillimeter galaxies (SMGs) at $z=1.4-4.1$ \citep{Ivison11, Bothwell13} are shown for the comparison. The fits of SMGs and SMGs together with (U)LIRGs \citep{Bothwell13} are also plotted.

Figure \ref{fig:SFE_fg} (a) shows CO(1--0) line luminosity $L'_{\rm CO(1-0)}$ and IR luminosity $L_{\rm IR}$ of TN J0924.
TN J0924 (all) shows that in this galaxy, the molecular gas is abundant comparing with objects which have similar $L_{\rm IR}$.
Also, TN J0924 (ABC) is an object with abundant molecular gas but low $L_{\rm IR}$.
While if we assuming that only clump A is a part of TN J0924, which is indicated by TN J0924 (H\&A), 
its high $L'_{\rm CO(1-0)}$ is reduced by a factor of 2.5 and becomes comparable with that of other HzRGs. 

On the other hand, TN J0924 (H) locates close to the fits of SMGs and (U)LIRGs \citep{Bothwell13}, consistent with being an ULIRG-like object with the upper limit of $L'_{\rm CO(1-0)}<1.7\times10^{10}$  $\rm \,K\,km\,s^{-1}pc^{2}$. 
In addition, regarding a $L_{\rm IR}\sim10^{12}$ $L_{\odot}$ object, its dust mass is estimated to be on the order of $10^{7} M_{\odot}$ \citep{Schreuber18}. Assuming a gas-to-dust ratio of 100, the molecular gas mass should be about $10^{9} M_{\odot}$, which is consistent with our constraint that the molecular gas mass at the host galaxy is $<1.3\times10^{10} M_{\odot}$, with $\alpha_{\rm CO} = 0.8$ $M_{\rm \odot}\rm\,(K\,km\,s^{-1}pc^{2})^{-1}$. 
As this upper limit is still higher than the best fits of SMGs and (U)LIRGs, it is consistent for the host galaxy to maintain its SFR of $110\pm20$ $M_{\odot}\,\rm yr^{-1}$.

Figure \ref{fig:SFE_fg} (b) shows star formation efficiency SFE $\equiv$ SFR/$M_{\rm H_{2}}$ and depletion time $t_{\rm depl}\equiv$ $M_{\rm H_{2}}$/SFR = SFE$^{-1}$ versus molecular gas fraction $f_{g}$. TN J0924 is shown in four sets as well. To estimate the upper limit of stellar masses of molecular gas clumps, we apply a simple estimation. The stellar mass $10^{11.1} M_{\odot}$ of the host galaxy is derived by SED fitting with its IRAC detection and other upper limits in the longer wavelengths \citep{DeBreuck10}.
Since only stellar population model is involved for SED fitting at IRAC wavelengths, we use the same IRAC images to simply apply a scaling estimation.
The $3\sigma$ upper limit of IRAC 3.6 $\mu$m photometry at clumps is one fourth of that at the host galaxy, so we estimate the $3\sigma$ upper limit of stellar mass as $3\times10^{10} M_{\odot}$. 
This upper limit is only applied on the molecular gas clumps. When the host galaxy is combined, we only consider the stellar mass of the host galaxy.

Here we mark TN J0924 (all) with $\alpha_{\rm CO} = 0.8$ $M_{\rm \odot}\rm\,(K\,km\,s^{-1}pc^{2})^{-1}$ for the comparison with other star-forming objects (SMGs and HzRGs). But we extend its error bar to cover the properties which are derived by applying $\alpha_{\rm CO} = 4.3$ $M_{\rm\odot}\rm\,(K\,km\,s^{-1}pc^{2})^{-1}$ on clump B and C.
Because the upper limits of SFR are only $<40$ $M_{\odot}$/yr, indicating that star formation in the clumps is not as active as ULIRGs, the conditions of interstellar medium (ISM) might be more similar to that in Milky-Way-like (MW) galaxies rather than ULIRGs.
If we adopt the MW conversion factor $\alpha_{\rm CO} = 4.3$ $M_{\rm \odot}\rm\,(K\,km\,s^{-1}pc^{2})^{-1}$, the molecular gas masses become five times larger than what we originally estimated. 
Therefore, we regard the above estimates on molecular gas mass as lower limits.
In this regard, the errors would be even larger if the $\alpha_{\rm CO}$ applied on clump A is higher than the assumed 0.8 $M_{\rm \odot}\rm\,(K\,km\,s^{-1}pc^{2})^{-1}$. And this would be the same case for TN J0924 (H\&A) as well.
As we expected, compared with other HzRGs, TN J0924 (all) is at a special position due to its high molecular gas fraction. While TN J0924 (H\&A) appears to be a typical radio galaxy with low molecular gas fraction and high SFE.

\subsection{Interpretations of clumps}

The intriguing spatial distribution of radio jet, host galaxy and multiple molecular gas clumps implies that TN J0924 is in a special phase of galaxy evolution. Our observation reveals an radio-CO alignment between the radio jet axis and the clump A, which is similar with several cases observed at lower redshifts \citep{Emonts14}. Note that there is a difference between TN J0924 and those in \cite{Emonts14}. In \cite{Emonts14}, the CO detection is found on the side where the radio source is the brightest. While in TN J0924, clump A is on the west side, where the component of radio continuum is weaker.
The physical origin of radio-CO alignment and off-center CO is considered as either outflow, merger, or metal enrichment \citep{Klamer04,Emonts14}. 
\citet{Emonts14} suggest that molecular gas reservoirs exist in the halo environment ($\sim50$ kpc scale) of HzRGs at $z\sim2$. Our result, three molecular gas clumps locate 12--33 kpc away from TN J0924, within its halo environment, is consistent with previous studies while at $z\sim5$.
Observations of multi-phase gas also suggest that the environment in the off-center CO is consistent with the ionization front of AGN \citep{Falkendal21}. But it is difficult for us to apply the photo-dissociated region model on TN J0924 without the information of multi-phase gas.

Considering that TN J0924 locates in an overdense region of Ly$\alpha$ emitters \citep{Venemans04} and Lyman break galaxies \citep{Overzier06}, merger event can possibly explain the exist of clumps. However, this can not explain the radio-CO alignment since the companion object does not necessarily align with the jet axis.
But certainly, the possibility of clump A being a merger can not be ruled out, merger event can be aligned with the jet axis by chance. Furthermore, if the jet axis is aligned with the major axis of the elliptical orbit of the merger, the probability of the alignment between the merger and the jet is likely to be higher.

Previous studies \citep{Klamer04,Emonts14} also suggest that as jet can induce the compression of gas, massive stars may primarily form along the jet. Carbon and oxygen produced by massive stars are thus more abundant along the jet axis. Therefore, CO are more abundant so that more detectable along the jet axis due to the jet-induced metal enrichment. 
However, numerical simulations which have been done by \citet{Mandal21} may provide a different perspective.
By combining the jet-induced turbulence and compression occurring in radio galaxy, \citet{Mandal21} show that the SFR is enhanced in the inner region of galaxy rather than in the region along the jet. 
However, observational studies show that the anisotropic metal enrichment is related to AGN jets \citep[e.g.,][]{Kirkpatrick11}. This may suggest that the jet-induced metal enrichment is still possible, but the origin of metal is not necessarily related to the jet-induced star formation.
Besides, \citet{Mandal21} show that jet-ISM interaction globally regulate the SFR of the galaxy. And the regulation is much effective at the outskirt than the inner region. The enhanced SFR at the inner region can exhaust the molecular gas at the host galaxy faster than at the outskirt. Thus, the molecular gas at the outskirt remains abundant due to the local low SFR.
This may explain the observed high SFR at the host galaxy where CO(1--0) is non-detected, and the off-center clump A which has no $HST$, IRAC and ALMA counterparts.
Whereas, this can not explain the observed radio-CO alignment either.

Clumps can also be interpreted as outflow from the AGN. Observations of X-ray cavities in brightest cluster galaxies suggest that the orientation of jet axis is caused by the precession. \citet{Dunn06} show that in the brightest cluster galaxy (BCG) of Perseus cluster, the precession with an opening angle of 50$^{\circ}$ has the time scale of about $3\times10^{7}$ years, comparable or shorter than the period of radio-loud activity \citep[$10^{7}$--$10^{8}$ years;][]{Bird08}. 
In TN J0924, each clump can possibly be the outflow in each period of radio-loud activity of the AGN, because the time scale of large precession is shorter than the period of radio-loud activity. 
Furthermore, if we assume the velocity of outflow is $\sim500$ $\rm km\,s^{-1}$, the clump can be 30 kpc away from the AGN in $\sim6\times10^{7}$ years. This is consistent with the time scale of the radio-loud activity and the distance between the observed clumps and the AGN.
At $z=0.596$, the BCG of the Phoenix cluster shows outflow of $\sim10^{10} M_{\odot}$, which is $\sim50\%$ of the total molecular gas mass in the galaxy \citep{Russell17}. This is different with the off-center CO but showing that the outflow of such massive amounts of molecular gas is possible.
However, unlike X-ray cavities in clusters appearing in pairs, the clumps in TN J0924 are only detected on one side of the AGN.

On the other hand, regarding the companion molecular gas reservoirs, 
\citet{Gullberg16a} suggest that this type of objects can follow the accretion flow associated with the growth of the cosmic web of dark matter, toward HzRGs. 
As TN J0924 is in an overdense region \citep{Venemans04,Overzier06}, clump B and C can be the objects which follow the accretion flow and move toward TN J0924.
Considering that if clump B and C will eventually fall and thus fuel the star formation activity, TN J0924 may move between the positions of TN J0924 (H\&A) and TN J0924 (all) on Figure \ref{fig:SFE_fg} (b) during its evolution.
Assuming that clump B and C will simply free fall onto the host galaxy, the estimation of free-fall time $t_{\rm ff}$ is following:
\begin{equation}
    t_{\rm ff}=\frac{\pi}{2}\frac{R^{3/2}}{\sqrt{2G\left(M+m\right)}},
\end{equation}
where $R$ is the distance between the point source mass $M$ and $m$, and $G$ is the gravitational constant. Assuming $M$ is the stellar gas mass of the host galaxy and $m$ is the molecular mass of clump B or C, the estimation of $t_{\rm ff}$ is about 0.2--0.3 Gyr. 
This timescale is comparable with the depletion time of the host galaxy alone with clump A (H\&A).
Following this assumption, it is possible that they will merge with TN J0924 and continually fuel the further star formation in the host galaxy.
Nevertheless, as the molecular gas will be exhausted out due to the star formation if the SFR remains constant, TN J0924 will eventually move toward the left top corner of Figure \ref{fig:SFE_fg} (b), following other HzRGs.

\section{conclusions} \label{sec:conclusions}

We have observed TN J0924--2201 with VLA at 19 GHz 
to obtain a spatially resolved CO(1--0) image of one of the most distant radio galaxies known. 
We present this new VLA K-band observations with archival ALMA Band-6 data.
The main results from this work are:

\begin{itemize}
    \item We detect CO(1--0) in TN J0924--2201 with a beam size of $2''.3\times1''.2$. Three clumps (Clump A, B, and C from north to south) are detected.
    \item Clump A is about $2''.0$ (12 kpc) offsetting to the west of the nucleus of the TN J0924-2201 host galaxy, and apparently being aligning with the radio jet axis, showing the radio-CO alignment. Clump B and C are about $3''.9$ (24 kpc) and $5''.4$ (33 kpc) offsetting to the south west and the south, respectively.
    \item The CO(1--0) line luminosity $L'_{\rm CO(1-0)}$ of clump A, B and C are
    $(4.7\pm0.7) \times10^{10}$ $\rm \,K\,km\,s^{-1}pc^{2}$, $(3.2\pm0.5) \times10^{10}$ $\rm \,K\,km\,s^{-1}pc^{2}$ and $(3.7\pm0.5) \times10^{10}$ $\rm \,K\,km\,s^{-1}pc^{2}$, respectively. 
    With the CO-to-H$_{2}$ conversion factor $\alpha_{\rm CO} = 0.8$ $M_{\rm \odot}\rm\,(K\,km\,s^{-1}pc^{2})^{-1}$,
    the corresponding molecular gas masses of clump A, B and C are $(3.7\pm0.6)\times10^{10}$ $M_{\rm \odot}$, $(2.5\pm0.4)\times10^{10}$ $M_{\rm \odot}$, and $(3.0\pm0.4)\times10^{10}$ $M_{\rm \odot}$, respectively.
    The summation of these molecular gas masses agrees well with the previously reported molecular gas mass from the spatially unresolved CO measurements by \citet{Klamer05}. 
    \item The ALMA 1.3 mm continuum image shows that the host galaxy of TN J0924 is dominated by the thermal dust emission and locates at the same position of the western component of radio continuum. By extrapolating the power law with $\alpha=-1.7$ , the non-thermal synchrotron emission only contributes about 2\% of the flux density observed at 1.3 mm.
    \item With the robust ALMA 1.3 mm continuum detection, we fit the modified blackbody with parameters $\beta=2.5$ and $T_{\rm dust}=50$ K. We therefore estimate the $L_{\rm IR}$ and the SFR of the host galaxy of TN J0924 as $(9.3\pm1.7)\times10^{11}$ $L_{\rm \odot}$ and $110\pm20$ $M_{\odot}\,\rm yr^{-1}$.
    At the host galaxy, we constrain on the $M_{\rm H_{2}}$ with $3\sigma$ upper limit of $1.3\times10^{10} M_{\rm \odot}$.
    \item At molecular gas clumps, by applying a simple scaling estimation with ALMA 1.3 mm and IRAC 3.6 $\mu$m continuum images, we constrain on $L_{\rm IR}$, SFR and $M_{*}$ of molecular gas clumps as $<3.5\times10^{11}L_{\odot}$, $<40$ $M_{\odot},\rm yr^{-1}$ and $<3\times10^{10}M_{\odot}$, respectively.
\end{itemize}

From these results, we demonstrated the properties of different combinations of the host galaxy and the clump A, B and C.
We also discussed on different scenarios about the interpretations of the phenomenon of off-center CO and radio-CO alignment.
Regarding the origin or the fate of the host galaxy and its companions, further observational data that can reveal their dynamic properties are crucial.

\acknowledgments

We thank the reviewer for the comprehensive and constructive comments.
K.L. acknowledges support from Japan-Taiwan Exchange Association.
F.E. is supported by JSPS KAKENHI grant No. JP17K14259.
M.I. acknowledges support from JSPS KAKENHI grant No. JP21K03632.
H.U. acknowledges support from JSPS KAKENHI grant (20H01953).
This study was supported by the JSPS KAKENHI Grant Number JP17H06130 and the NAOJ ALMA Scientific Research Number 2017-06B.

The National Radio Astronomy Observatory is a facility of the National Science Foundation operated under cooperative agreement by Associated Universities, Inc.
This paper makes use of the following ALMA data: ADS/JAO.ALMA\#2013.1.00039.S. ALMA is a partnership of ESO (representing its member states), NSF (USA) and NINS (Japan), together with NRC (Canada), MOST and ASIAA (Taiwan), and KASI (Republic of Korea), in cooperation with the Republic of Chile. The Joint ALMA Observatory is operated by ESO, AUI/NRAO and NAOJ.

\bibliography{Mybib}{}
\bibliographystyle{aasjournal}

\end{document}